\documentclass[12pt] {article}
\usepackage{graphics}
\usepackage{epsfig}

\catcode`@=12 \topmargin -0.5in \evensidemargin 0.0in \oddsidemargin
0.0in \textheight 8.5in \textwidth 6.5in

\newcommand{\beq}{\begin{equation}}
\newcommand{\eeq}{\end{equation}}

\newcommand{\bea}{\begin{eqnarray}}
\newcommand{\eea}{\end{eqnarray}}

\begin{document}
\thispagestyle{empty}
\begin{flushright}
UMD-PP-06-004\\
OSU-HEP-06-02\\
\end{flushright}
\vspace{0.5in}
\begin{center}
{\LARGE \bf Post--Sphaleron Baryogenesis\\} \vspace{1.3in} {\bf
K.S. Babu$^1$, R.N. Mohapatra$^2$ and S. Nasri$^2$\\}
\vspace{0.2in} {\sl $^1$ Department of Physics, Oklahoma State
University, Stillwater, Oklahoma 74078, USA\\} {\sl $^2$
Department of Physics, University of Maryland, College Park, MD
20742, USA\\} \vspace{1in}
\end{center}

\begin{abstract}

We present a new mechanism for generating the baryon asymmetry of
the universe directly in the decay of a singlet scalar field $S_r$
with a weak scale mass and a high dimensional baryon number
violating coupling.  Unlike most currently popular models, this
mechanism, which becomes effective after the electroweak phase
transition, does not rely on the sphalerons for inducing a nonzero
baryon number. CP asymmetry in $S_r$ decay arises through loop
diagrams involving the exchange of  $W^\pm$ gauge bosons, and is
suppressed by light quark masses, leading naturally to a value of
$\eta_B \sim 10^{-10}$. We show that the simplest realization of
this mechanism, which uses a six quark $\Delta B=2$ operator,
predicts colored scalars accessible to the LHC, and
neutron--antineutron oscillation within reach of the next
generation experiments.

\end{abstract}

\newpage
\baselineskip 24pt



\section{Introduction}

Recent developments in particle physics have had profound impact
on cosmology.  One of the most far--reaching consequences has been
the possibility that new interactions beyond the standard model
can explain the origin of matter--antimatter asymmetry of the
universe as a dynamical phenomenon. There are currently several
attractive scenarios which achieve this, the two most widely
discussed ones being (i) baryogenesis via leptogenesis
\cite{lepto}, which is connected to the seesaw mechanism and
neutrino masses, and (ii) weak scale baryogenesis \cite{nelson},
which involves supersymmetric or multi--Higgs extensions of the
standard model. Both these proposals depend crucially on the
properties of the elctroweak sphaleron \cite{kuzmin} which serves
as the source of $B$ violation. Since the nature of new physics
beyond the standard model remains unknown presently, it is
important to explore alternative mechanisms that can explain the
matter--antimatter asymmetry while yielding testable consequences.
In this letter we suggest and explore one such alternative.

The salient feature of our proposal is that baryogenesis occurs
via the direct decay of a scalar boson $S_r$ having a weak scale
mass and a high dimensional baryon violating coupling.  $S_r$ is
the real part of a baryon number carrying complex scalar $S$,
which acquires a vacuum expectation value (vev). The decays $S_r
\rightarrow 6q$ and $S_r \rightarrow 6 \overline{q}$ will then be
allowed, providing the source for $B$ asymmetry. These decays
occur when the temperature of the universe is $T \sim 0.1-100$
GeV. By this time the electroweak sphalerons have gone out of
thermal equilibrium, and thus play no role in the $B$ asymmetry
generation. We call this mechanism ``post--sphaleron
baryogenesis". The three Sakharov conditions for successful
baryogenesis \cite{sakharov} are satisfied rather easily in our
scheme.  The high dimensionality of the $B$ violating coupling of
$S_r$ to the quark fields allows the $\Delta B\neq 0$ decays to go
out of equilibrium at weak scale temperatures.  CP violation
occurs in the decay via loop diagrams involving the exchange of
the standard model $W^\pm$ gauge bosons. This amplitude has
sufficient light quark mass suppression to explain naturally the
observed (small) value of the baryon to photon ratio $\eta_B \sim
10^{-10}$. The simplest realization of our mechanism involves
interactions that violate $B$ by two units and therefore gives
rise to neutron--antineutron oscillations. We find that the
successful implementation of our mechanism sets an upper limit on
the transition time for $N \leftrightarrow \bar{N}$ oscillation
bringing it to within the realm of observability. This connection
provides a strong motivation for improved searches for $N
\leftrightarrow \bar{N}$ oscillation \cite{kamyshkov}.

The connection with $N \leftrightarrow \bar{N}$ oscillation can be
understood as follows. Let us consider an interaction of the form
$S{\cal O}_{\Delta B}$, where $S$ is a standard model singlet
complex scalar field (with $S_r$ denoting its real part) and
${\cal O}_{\Delta B}$ is the baryon number violating operator in
question. This interaction will lead to baryon number violation if
$\left\langle S \right\rangle \neq 0$. Suppose the mass dimension
of the operator $S {\cal O}_{\Delta B}$ is $M^{-n}$ with $n$
positive. The higher the value of $n$ for an operator ${\cal
O}_{\Delta B}$, the lower the mass scale allowed by the existing
limits on baryon violation. Since the rate of these $\Delta B\neq
0$ interactions in early universe goes like $M^{-2n}$, the higher
the value of $n$, the easier it is to satisfy the
out-of-equilibrium condition at a lower temperature (multi GeV
range). Clearly the operator leading to $B-L$ conserving proton
decay mode cannot be useful for us, since present experimental
limits on proton lifetime imply that this operator should go out
of equilibrium at temperatures of order $10^{14}-10^{15}$ GeV. On
the other hand, for a process like $N \leftrightarrow \bar{N}$
oscillation \cite{kuzmin1,glashow,marshak}, present experimental
lower limits on the oscillation time $\tau_{N-\bar{N}}$
\cite{milla,nnbar} allow the mass $M$ appearing in the operator
$(fgh)u^cd^cd^cu^cd^cd^c/M^5$ to be in the multi--TeV range (for
the first family Yukawa couplings $f \sim g \sim h \sim 10^{-3}$).
The out-of-equilibrium temperature for the processes
$S_r\rightarrow 6 q$ and $S_r \rightarrow 6 \overline{q}$ is then
allowed to be below the sphaleron decoupling temperature of about
100 GeV. We will illustrate how post--sphaleron baryogenesis works
using the $\Delta B=2$ process, although the mechanism applies
more generally.  (The $\Delta B = 2$ operator involving
left--handed quark doublet fields, $QQQQQQH^*H^*/M^7$, has
additional Higgs fields and thus a higher dimensionality.)

The high dimensional $\Delta B = 2$ couplings of $S$ are obtained
by integrating out colored scalar fields.  These colored scalars
cannot be much heavier than about a TeV, or else the induced
$\eta_B$, consistent with nucleosynthesis limits, will be too
small. The prospects for discovering such baryon number carrying
colored scalars at the Large Hadron Collider are quite promising.

An attempt to generate baryon asymmetry at a temperature of order
MeV via the decay of a heavy ($\sim 50$ TeV) gravitino within
supergravity was proposed in Ref. \cite{cline}.  Such a large
gravitino mass would however require fine-tuning to solve the
hierarchy problem. Another scenario \cite{hall} invokes the decay
of the inflaton into squarks, with their subsequent decay
producing baryon asymmetry.  This mechanism requires that the
reheating temperature be less than a GeV in order for the
scattering and inverse decays not to wash out the asymmetry.  The
model presented here differs from these earlier attempts in two
crucial ways:  (i) There is a strong link between baryon asymmetry
and $N \leftrightarrow \overline{N}$ oscillation, and (ii) the
mechanism of inducing CP asymmetry via the standard model $W^\pm$
loops which leads naturally to a value of  $\eta_B \sim 10^{-10}$
due to light quark mass suppression is entirely new.

\section{ Light diquarks and observable $N \leftrightarrow \bar{N}$ oscillation}

To illustrate our mechanism for post--sphaleron baryogenesis, we
consider a generic TeV scale model that gives rise to the higher
dimensional operator  for ${N\leftrightarrow \bar{N}}$
oscillation. It consists of the following color sextet, $SU(2)_L$
singlet scalar bosons $(X,Y,Z)$ with hypercharge $-\frac{4}{3},
+\frac{8}{3}, +\frac{2}{3}$ respectively that couple to the
right--handed quarks.\footnote{Color sextet fields are preferred
over color triplets, since the sextets do not mediate proton
decay.} In addition, there is a complex scalar field $S$ which is
a singlet of the standard mode with mass in the 100 GeV range.
With this field content one can write down the following standard
model invariant interaction Lagrangian:\footnote{An additional
term $g'_{ij}Z(u^c_id^c_j-u^c_jd^c_i)$ with $g'_{ij} = -g'_{ji}$
is also allowed by the standard model gauge symmetry, but this
term is forbidden when the model is embedded minimally into a
left--right symmetric framework. We do not keep this term
explicitly here, its inclusion is however straightforward.}
\begin{equation}
{\cal L}_I = {h_{ij} \over 2}X d^c_id^c_j + {f_{ij} \over 2}
Yu^c_iu^c_j + {g_{ij} \over 2} Z (u^c_id^c_j+u^c_jd^c_i)
+~{\lambda_1 \over 2} S X^2Y~+~{\lambda_2 \over 2} SXZ^2 + h.c.
\end{equation}
If the scalar field $S$ which has $B=2$ is given a vacuum
expectation value, cubic scalar field couplings of the type $X^2Y$
that break baryon number by two units will be induced.  In turn it
will lead to ${N\leftrightarrow \bar{N}}$ oscillation via the
diagram of Fig. 1  with $S_r$ replaced by $\left\langle S
\right\rangle$ \cite{marshak}.

We note that not all of the $(X,Y,Z)$ fields are needed for $B$
violation and $N \leftrightarrow \overline{N}$ oscillation.  $(X,Y)$
or $(X,Z)$ fields will do.  We will focus more on these minimal
versions in our computation of baryon asymmetry, while for
generality we keep all three fields.

To see the constraints on the parameters of the theory, we note
that the present limits on $\tau_{N-\bar{N}}\geq 10^8$ sec.
implies that the strength $G_{N-\bar{N}}$ of the $\Delta B= 2$
transition is $\leq 10^{-28}$ GeV$^{-5}$. From Fig. 1, we conclude
that
\begin{eqnarray}
G_{N-\bar{N}}\simeq \frac{\lambda_1 \left\langle S \right \rangle
h^2_{11} f_{11}}{M^2_YM^4_X}~+~\frac{\lambda_2 \left\langle S
\right \rangle h_{11}g^2_{11}}{M^2_XM^4_Z}\leq 10^{-28}~ {\rm
GeV}^{-5}.
\end{eqnarray}
For $\lambda_{1,2} \sim 1,~h_{11} \sim f_{11} \sim g_{11} \sim
10^{-3}$, we find $\left \langle S \right \rangle \sim
M_{X,Y,Z}\simeq 1$ TeV is allowed. In our discussion, we will stay
close to this range of parameters and see how one can understand
the baryon asymmetry of the universe. In fact, we will see that
the masses of $X,Y,Z$ cannot be much larger than a TeV for
successful baryogenesis. Note that the couplings $(f,g,h)_{ij}$ to
the second and third generation fermions could be larger.

Other constraints can come from low energy observations such as
bounds on flavor changing hadronic processes such as $K-\bar{K}$,
$D-\bar{D}$ transition etc. This of course depends on any possible
mixings between the right handed quark fields, on which we do not
have any apriori information. If we make the simplest assumption
dictated by the left--right symmetric theories that the left and
the right--handed mixings are equal, then the strongest
constraints come from $K-\bar{K}$ transition which imply that for
$h_{11} \sim 10^{-3}$, $M_X\geq 1$ TeV, which is consistent with
our choice of parameters dictated by observability of $N
\leftrightarrow \bar{N}$ transition.

The model of Eq. (1) is embeddable into an $SU(2)_L\times
SU(2)_R\times SU(4)_c$ framework where the quarks and leptons
transform as $\psi:({\bf 2,1,4})~\oplus~ \psi^c:({\bf
1,2},\bar{\bf 4})$ representations and the Higgs fields $X,Y,Z,S$
are part of the $\Delta^c:({\bf 1,3,10})$ multiplet. In fact the
$S$ field corresponds to the $\Delta^c_{\nu^c\nu^c}$ component
that acquires a vev and breaks $B-L$ by two units. In this paper,
we will not discuss the full set of constraints that arises in
this embedding but rather simply work within the scalar field
model described in Eq. (1). All our conclusions below apply to the
$SU(2)_L \times SU(2)_R \times SU(4)_c$  model as well.  While we
take baryon number as part of the gauge symmetry, the mechanism of
$B$ asymmetry generation also works if $B$ is a spontaneously
broken global symmetry as in Ref. \cite{barbieri}.

\section{Origin of matter}

 Before proceeding to the discussion of
how baryon asymmetry arises in this model, let us first consider the
effect of the new interactions in Eq. (1) on any pre-existing baryon
asymmetry. For this purpose, we assume the following mass hierarchy
between the $S$ field and the $(X,Y,Z)$ fields: $M_S\sim 100$ GeV
$\ll M_{X,Y,Z}\sim$ TeV. For $T\geq M_{X,Y,Z}$, the $\Delta B=2$
interaction rates scale like $T$ and are in equilibrium at least
down to $T\simeq M_{X,Y,Z}$. They will therefore erase any
pre-existing baryon asymmetry. They remain in equilibrium down to
the temperature $T_*$ determined by the inequality:
\begin{equation}
{1 \over (2\pi)^9} {\lambda_2^2h^2 g^4 T^{13} \over
M^{12}_{X,Z}}\leq \frac{g^{1/2}_*T^2}{M_{Pl}}~.
\end{equation}
Here $h$ and $g$ refer to the largest of $h_{ij}$ or $g_{ij}$
($i,j$ are family indices). For $h,g$ in the range of $0.1-1$,
this leads to $T_*\simeq (0.6 - 0.2) M_{X,Z}$.

The singlet field $S$ will play a key role in the generation of
baryon asymmetry. We assume that $\left\langle S\right\rangle \sim
M_{X}$ and $M_{S_r}\sim 10^2$ GeV, where $S_r$ is the real part of
the $S$ field after its vev is subtracted. $S_r$ can then decay
into final states with $B=\pm 2$, viz., $S_r \rightarrow 6 q$ and
$S_r \rightarrow 6 \overline{q}$, inducing a net baryon asymmetry.

On the way to calculating the baryon asymmetry, let us first
discuss the out of equilibrium condition. As the temperature of
the universe falls below the masses of the $X,Y,Z$ particles, the
annihilation processes $X\bar{X}\rightarrow d^c\bar{d^c}$ (and
analogous processes for $Y$ and $Z$) remain in equilibrium. As a
result, the number density of $X,Y,Z$ particles gets depleted and
only the $S$ particle survives along with the usual standard model
particles. The primary decay modes of $S_r$ are $S_r\rightarrow
u^cd^cd^cu^cd^cd^c$ and $S_r \rightarrow \bar{u}^c
\bar{d}^c\bar{d}^c\bar{u}^c \bar{d}^c\bar{d}^c$. There could be
other decay modes which depend on the details of the model. Those
can be made negligible by choice of parameters which do not affect
our discussions of $N \leftrightarrow \bar{N}$ oscillation and
baryogenesis. We will discuss them later in the paper.  For $T\geq
M_{S_r}$, the decay rate of $S_r$ is given by the left--hand side
of Eq. (3). This decay goes out of equilibrium around $T_* \sim
0.4 M_X$, or around 500 GeV. Below this temperature the decay rate
of $S_r$ falls very rapidly as the temperature cools. However as
soon as $T\leq M_{S_r}$, the decay rate becomes a constant while
the expansion rate of the universe slows down. So at a temperature
$T_d$, $S$ will start to decay where $T_d$ is given by
\begin{equation}
T_d \simeq \left[\frac{18 P \lambda_2^2h^2
g^4M_{P\ell}M_{S_r}^{13}}{(2\pi)^9 1.66
g_*^{1/2}(6M_X)^{12}}\right]^{1/2}~.
\end{equation}
This is obtained by equating the decay rate of $S_r$ to the
expansion rate of the universe.  In Eq. (4) the factor 18 is a
color factor, $h^2 = {\rm Tr}({h^\dagger h})$, etc, while $P$ is a
phase space factor, which we have computed for the six body decay
via Monte Carlo methods and found $P \simeq 2.05$. The
corresponding epoch must be above that of big bang
nucleosynthesis. This puts a constraint on the parameters of the
model. For instance, for $M_S\sim 200 $ GeV and $M_X\sim $ TeV, we
get $T_d\sim 40 $ MeV (for $g \sim h \sim 1$). We will conduct the
rest of the discussion with this set of parameters as a
representative set. Note that $M_X$ cannot be much larger than
about 1 TeV, otherwise $T_d$ will be below few MeV, affecting big
bang nucleosynthesis significantly.  Note also that the at least
some of the couplings in $h$ and $g$ should be of order one. This
would imply that the first family couplings should be of order
$(10^{-3}-10^{-4})$ from naturalness ($h_{11} \sim V_{td}^2
h_{33}$ etc), making $N \leftrightarrow \overline{N}$ oscillation
accessible to next generation experiments.

We now proceed to calculate the baryon asymmetry in this model. It
is well known that baryon asymmetry can arise only via the
interference of a tree diagram with a one loop diagram which has
an absorptive part. The tree diagrams are clearly the one where
$S_r\rightarrow 6 q$ and $S_r \rightarrow 6 \overline{q}$. There
are however two classes of loop diagrams that can contribute to
baryon asymmetry: one where the loop involves the same fields
$X,Y$ and $Z$ as in Fig. $2$ (there is a second loop diagram of
this type with $(X,Z)$ fields inside the loop), and a second one
involving $W^\pm$ gauge boson exchange as shown in Fig. $3$. In
the $(X,Z)$ model and in the ($X,Y)$ model, only the latter
contribution exists (the former trace being real). So we focus on
that latter, which involves only standard model physics at this
scale and has the advantage that it involves less number of
arbitrary parameters.

\begin{figure}
\centering \epsfysize=2.5in \epsffile{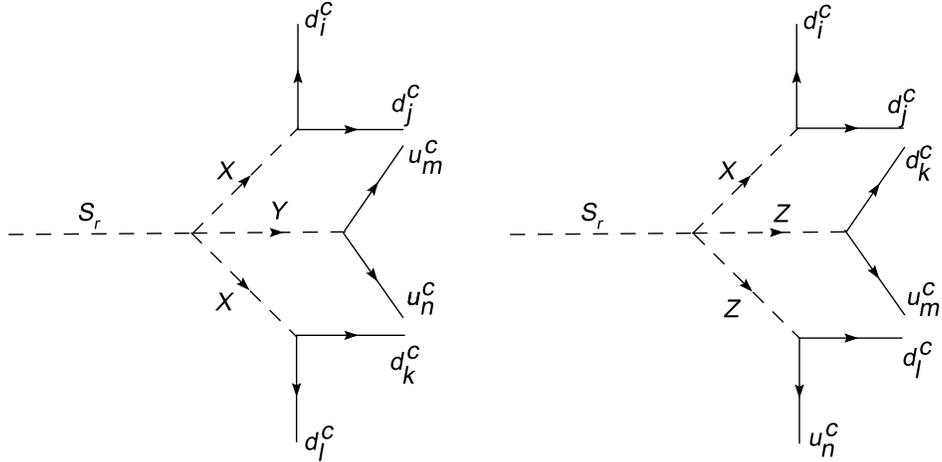} \caption{Tree level
diagrams contributing to $S_r$ decays into 6 anti-quarks. There are
other diagrams where $S_r$ decays into 6 quarks, obtained from the
above by reversing the arrows of the quark fields.}
\end{figure}

\begin{figure}
\centering \epsfysize=2.5in \epsffile{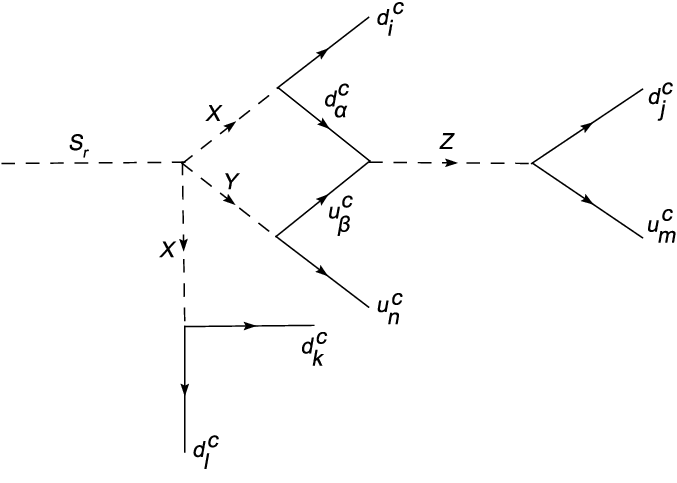} \caption{Diamond
loop diagram in the $(X,Y,Z)$ model.}
\end{figure}

We summarize the results of our calculations for the $W^\pm$
exchange diagrams. If one of the external up--type quarks is the top
quark, the corresponding quark line receives a wave function
correction via $W^\pm$ gauge boson exchange. The baryon asymmetry
from this diagram is found to be

\begin{equation}
{\epsilon_B^{\rm wave} \over {\rm Br}} \simeq -{3  \alpha_2 \over
8} \left(1+{m_W^4 \over m_t^4}\right) {{\rm Im}\left[V^*
\hat{M}_d^2 V^T \hat{M_u} g g^{\dagger}\right]_{33} \over m_t
m_W^2 (g g^\dagger)_{33}}
\end{equation}
where $\hat{M}_u = diag(m_u,~m_c,~m_t)$, $\hat{M}_d =
diag(m_d,~m_s,~m_b)$ and $V$ is the CKM matrix.  Br stands for the
branching ratio of $S_r$ into $6q + 6 \overline{q}$.

The vertex correction via the $W$ boson exchange gives an asymmetry
given by
\begin{equation}
{\epsilon_B^{\rm vertex} \over {\rm Br}} \simeq - {\alpha_2 \over
4} {{\rm ImTr}[g^T \hat{M}_u V g^\dagger V^* \hat{M}_d] \over {\rm
Tr}(g^\dagger g) }~.
\end{equation}
Here we have assumed that $M_{S_r} \gg m_t$.  In the limit where
$m_{S_r} \ll m_W$, we have the same asymmetry as in Eq. (6), but
with a factor of (-1/4) multiplying it.  Of course in this case,
decays involving final state top quark are disallowed, which is to
be implemented by removing the top quark contribution in the trace
of Eq. (6).\footnote{These $W^\pm$ loops do not conflict with the
theorem of Ref. \cite{weinberg} which states that no baryon
asymmetry can be induced by dressing a $\Delta B = 1$ vertex by
baryon number conserving interactions.  Since $S_r$ field has no
definite baryon number, owing to $\left\langle S \right\rangle
\neq 0$, the theorem is not applicable in our case.}

\begin{figure}
\centering \epsfysize=2.0in \epsffile{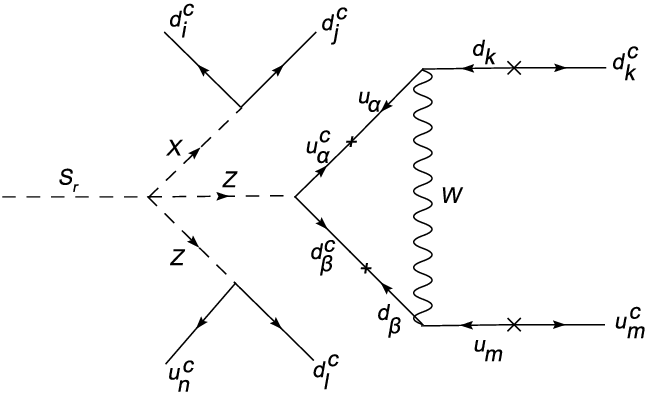} \caption{One loop
vertex correction diagram for the $B$-violating decay
$S_r\rightarrow 6q$.  There are also wave function corrections
involving the exchange of $W^\pm$ gauge boson.}
\end{figure}

It is interesting to note that in this mechanism, there is a
natural explanation of the observed baryon asymmetry $\eta_B \sim
10^{-10}$.  It follows from the light quark mass and mixing angle
suppression.  As an example, consider the following choice of
parameters: $m_c(m_c) = 1.27\;GeV$, $m_b(m_b) = 4.25\;GeV$, $m_t=
 174\;GeV$, $V_{cb} \simeq 0.04$, $M_S=200\;GeV$ and $|g_{33}| \simeq |g_{23}|\sim
 1$, with smaller values of $g_{1i}$.  We find $\epsilon_B \sim 10^{-8}$ in this case
 from Eq. (5).  The corresponding value from Eq. (6) is an order
 of magnitude larger, for the same input parameters.

There is a further dilution of the baryon asymmetry arising from
the fact that $T_d \ll M_{S_r}$ since the decay of $S_r$ also
releases entropy into the universe. In this case the baryon
asymmetry reads
\begin{equation}
\eta_B \simeq \epsilon_B\frac{T_d}{M_{S_r}}~.
\end{equation}
In order that this dilution effect is not excessive, there must be
a lower limit on the ratio $T_d/M_{S_r}$. From our estimate above
we require that $T_d/M_{S_r} \geq 0.01$. Since the decay rate of
the $S_r$ boson depends inversely as a high power of $M_{X,Y}$,
higher $X,Y$ bosons would imply that $\Gamma_S \sim H$ is
satisfied at a lower temperature and hence give a lower
$T_d/M_{S_r}$. In Fig. $4$ we have plotted $M_{X,Y}$ vs $M_{S_r}$
which gives the right amount of baryon asymmetry which is
consistent with the demand that the decay of $S_r$ occurs before
the QCD phase transition (i.e $T_d \geq 200\;MeV$). Using the
effective coupling $\bar{\lambda} \equiv
(\lambda_1h_{11}^2f_{11})^{1/4} \sim (\lambda_2hg_{11}^2)^{1/4}$
to be  of order $10^{-4}$ implies that the $10^{9}\;sec\leq
\tau_{N-\bar{N}} \leq 10^{11}\;sec$ for $M_{S_r}$ in range of
$\simeq 100 - 300\; GeV$.

\begin{figure}
\centering \epsfysize=3in \epsffile{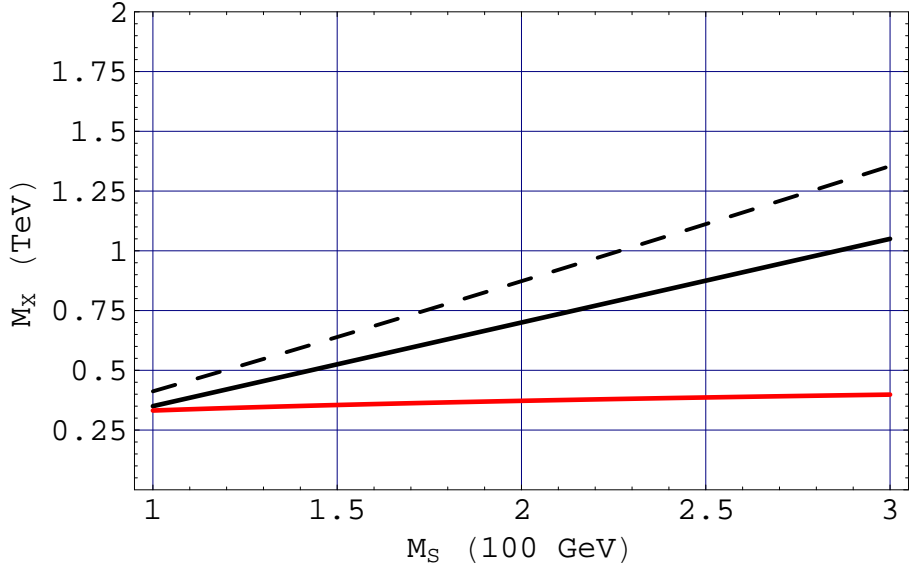} \caption{The
allowed range of $M_X$ and $M_S$ needed to generate the baryon
asymmetry (along the black curve), decay temperature above
$200\;MeV$ (points below the dashed curve) and $\tau_{N\bar{N}} \geq
10^8\;sec$ with $\bar{\lambda} = 10^{-4}$ (points above the red
curve).}
\end{figure}

We now comment briefly on some other aspects of the model:

(i) If the ($X,Y,Z)$ scalars are all present, the loop diagram of
Fig. 2 will contribute to baryon asymmetry.  Since there are two
diagrams of the type shown in Fig. 2 the relevant trace has an
imaginary piece.  The asymmetry will have a suppression factor
$M_{S_r}^2/M_X^2$, in addition to a loop factor and the Yukawa
suppression.  Although not very predictive, this model can  yield
adequate baryon asymmetry.

 (ii) The singlet field $S$ can have a
renormalizable interaction with the standard model Higgs doublet
of the form $\lambda_S S^{\dagger}SH^{\dagger}H$. After the fields
$S$ and $H$ acquire vacuum expectation values, the Re $S$ and the
SM Higgs field can mix with each other opening up new decay
channels for the $S_r$ field such as $S_r\rightarrow b\bar{b}$
etc. We estimate this decay width to be
\begin{eqnarray}
\Gamma(S_r\rightarrow b\bar{b})\sim \frac{3 \beta^2m^2_bM_S}{4\pi
M^2_W}
\end{eqnarray}
where $\beta$ is the $S_r-h$ mixing angle.
 There are two constraints on this decay mode in order for
our scenario to work: first, this could contribute dominantly to
the $S_r$ decay width thereby diluting the baryon asymmetry.
Second, this model should be out of equilibrium at $T=M_S$. If the
model is non-supersymmetric, these two conditions are satisfied if
$\lambda_S\leq 10^{-6}$. This coupling is automatically forbidden
if the model is supersymmetric.

(iii) The present considerations could be easily extended to
include supersymmetry as part of the new physics beyond the
standard model. The $SX^2Y$ and $SXZZ$ interactions in this case
are nonrenormalizable \cite{dutta}. However, in this case we also
expect mass terms in the superpotential of the form $M_SS\bar{S}$
so that the effective four scalar interaction responsible for
baryogenesis is in the same form as discussed above.

(iv) Our theory is also testable in collider experiments such as
the LHC since we have colored diquark scalar fields with masses in
the TeV range. In a $p\bar{p}$ collision one could produce the
$X,Y,Z$ bosons either in pairs via the process
$q\bar{q}\rightarrow X\bar{X}$ or singly via the process $q
+g\rightarrow X+\bar{q}$. In the first case the signal would be a
four jet final state whereas in the second case, it would be three
jet final states. It would therefore be important to search for
such final states at LHC.  One distinguishing feature of these
bosons is that they carry baryon number, which may be testable in
the decays of these bosons into top quark and bottom quarks.

\section{Conclusion}

In conclusion, we have presented a new mechanism for baryogenesis
which does not rely on the electroweak sphalerons but rather
directly produces matter--antimatter asymmetry using higher
dimensional baryon violating couplings of a scalar field. The
baryon asymmetry is produced at the weak scale. This mechanism can
be tested by searches for baryon violating processes such as
neutron antineutron oscillation, as well as by the discovery of
colored scalars at the LHC.

S.N would like to thank Mark Trodden for helpful comments and
discussions .The work of KSB is supported by DOE grant Nos.
DE-FG02-04ER46140 and DE-FG02-04ER41306, and that of RNM and SN is
supported by the National Science Foundation grant No. Phy-0354401.

\end{document}